\documentclass[5p,twocolumn]{elsarticle}

\usepackage{amssymb}

\journal{Journal of Supercritical Fluids}

\begin{document}

\begin{frontmatter}



\title{Large-area silica aerogel for use as Cherenkov radiators with high refractive index, developed by supercritical carbon dioxide drying}


\author[First]{Makoto Tabata\corref{cor1}}
\ead{makoto@hepburn.s.chiba-u.ac.jp}
\cortext[cor1]{Corresponding author.} 
\author[Second]{Ichiro Adachi}
\author[Third]{Yoshikiyo Hatakeyama}
\author[First]{Hideyuki Kawai}
\author[Third]{\\Takeshi Morita}
\author[Fourth]{Takayuki Sumiyoshi}

\address[First]{Department of Physics, Chiba University, Chiba 263-8522, Japan}
\address[Second]{Institute of Particle and Nuclear Studies (IPNS), High Energy Accelerator Research Organization (KEK), Tsukuba 305-0801, Japan}
\address[Third]{Graduate School of Advanced Integration Science, Chiba University, Chiba 263-8522, Japan}
\address[Fourth]{Department of Physics, Tokyo Metropolitan University, Hachioji 192-0397, Japan}

\begin{abstract}
This study presents the development of large-area (18 $\times $ 18 $\times $ 2 cm$^3$), high refractive index ($n \sim $1.05) hydrophobic silica aerogel tiles for use as Cherenkov radiators. These transparent aerogel tiles will be installed in a Cherenkov detector for the next-generation accelerator-based particle physics experiment Belle II, to be performed at the High Energy Accelerator Research Organization (KEK) in Japan. Cracking has been eliminated from the prototype aerogel tiles by improving the supercritical carbon dioxide (scCO$_2$) extraction procedure when drying the wet gel tiles. Finally, a method of mass-producing aerogel tiles for the actual detector was established. It was confirmed that the experimentally manufactured aerogel tiles meet the required optical and hydrophobic characteristics and have a uniform tile density.
\end{abstract}

\begin{keyword}
Silica aerogel \sep Supercritical drying \sep Sol--gel polymerization \sep Refractive index \sep Cherenkov radiator \sep Belle II

\end{keyword}

\end{frontmatter}



\section{Introduction}
\label{1}

Silica aerogel was among the first aerogel materials to be created from wet gel via supercritical drying \cite{cite1} that comprises three-dimensional networks of silicon dioxide (SiO$_2$) clusters and air-filled open pores of nanometer-scale dimensions. Silica aerogel (hereafter referred to simply as \textit{aerogel}) is essentially transparent under visible light; however, its transparency depends on its mesoporous nanostructure, which is controlled mainly by sol--gel polymerization in the wet-gel synthesis step. During this step, the bulk density of the aerogel can be tuned by varying the ratio of the silica precursor to the diluent solvent (e.g., 0.02--0.5 g/cm$^3$ \cite{cite2}). Besides ultralow-density aerogel (0.01 g/cm$^3$ or less) \cite{cite3,cite4}, researchers have prepared (ultra)high-density aerogel (0.2--1.0 g/cm$^3$) by applying the pin-drying technique to the synthesized wet gel \cite{cite4}. Native aerogel is hydrophilic, but can be rendered hydrophobic by modification of the silanol groups (Si--OH) on the surface of the silica clusters \cite{cite2,cite5}. Using recent techniques, hydrophobic, transparent aerogel blocks can be fabricated over a wide density range (0.01--1.0 g/cm$^3$) \cite{cite4}. The refractive index ($n$) of the aerogel is empirically proportional to its density ($\rho $):
\begin{equation}
\label{eq:eq1}
n(\lambda )-1=k(\lambda )\rho,
\end{equation}
where $k$ is a coefficient that depends on the light wavelength ($\lambda $) \cite{cite2}. This expression is theoretically derived from the Lorentz--Lorenz formula (e.g., \cite{cite6}).

\begin{figure}[t]
\centering 
\includegraphics[width=0.45\textwidth,keepaspectratio]{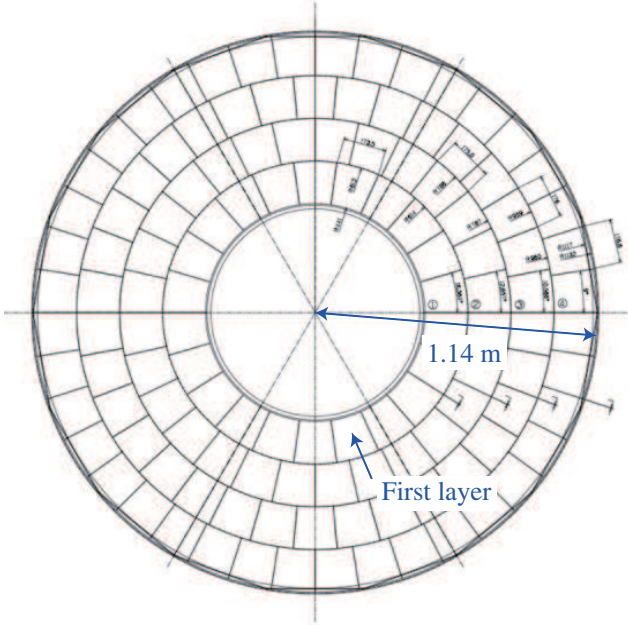}
\caption{CAD drawing of the planned A-RICH radiator tiling in the cylindrical forward end cap of the Belle II detector. The outer radius is 1.14 m. The tiling area comprises four concentric layers; the first layer is indicated by the arrow.}
\label{fig:fig1}
\end{figure}

Over three decades, we have been developing transparent silica aerogel with a wide range of refractive indices for use in scientific instruments, especially in Cherenkov radiators \cite{cite2,cite4,cite7,cite8,cite9,cite10,cite11,cite12}. One of our specific targets is the Belle II experiment, a next-generation \textit{B} factory experiment that will employ the SuperKEKB electron--positron collider at the High Energy Accelerator Research Organization (KEK) in Tsukuba, Japan \cite{cite13}. The Belle II will succeed the Belle experiment. From 1999 to 2010, Belle contributed remarkably to precise tests of the Standard Model of particle physics by measuring the decay modes of \textit{B} mesons generated by the annihilation of electron--positron pairs \cite{cite14}. One of the particle identification devices in the Belle II detector system (under construction) is an aerogel-based ring-imaging Cherenkov (A-RICH) counter. This counter is installed at the forward end cap region and requires a large number of transparent silica aerogel tiles with different refractive indices as the Cherenkov radiator \cite{cite15}. Figure \ref{fig:fig1} shows the planned aerogel radiator tiling of the A-RICH counter \cite{cite16}. In this tiling plan, a cylindrical area of 3.5 m$^2$ is covered with two layers of large-area aerogel tiles segmented into 124 blocks, each of thickness 2 cm (i.e., a total of 248 tiles). The fan-shaped aerogel segments will be trimmed from square aerogel tiles of dimensions 18 $\times$ 18 $\times $ 2 cm$^3$ using a water jet cutter. The final tile dimensions were determined from the results of the present study.

The goal of the present study is to reduce the adjacent Cherenkov radiator boundaries by minimizing the number of tiles that fill the radiator plane, which is vital for achieving high detector performance over the broad end cap area. This is equivalent to maximizing the tile cross-section of the 2 cm-thick aerogel radiator. Thus far, aerogel tiles with dimensions of up to 15 $\times $ 15 $\times $ 2 cm$^3$ have been fabricated \cite{cite10}, but the crack-free production yield was unstable (e.g., 50--80\%). Mass production of larger, crack-free aerogel tiles with the same high optical characteristics of small size aerogel tiles presents a major challenge. The optical requirements are the apparent refractive index, the transmission length defined in Section \ref{5-2-1}, and uniform internal tile density (i.e., refractive index uniformity). Specifically, the refractive indices must be 1.045 $\pm $ 0.002, 1.050 $\pm $ 0.002, and 1.055 $\pm $ 0.002, and the acceptable transmission lengths (measured at $\lambda $ = 400 nm) are 45, 40, and 35 mm or more, respectively. The transversal uniformity of each large-area aerogel tile should satisfy $|\delta (n - 1)/(n - 1)| < 4\%$. Moreover, the crack-free yield should ideally exceed 80\%. Only crack-free tiles will be used in the actual detector. Details of the requirements are provided in \ref{app}. Thus, the present study aims to fabricate large-area, crack-free, hydrophobic aerogel tiles with $n$ $\sim $1.05, corresponding to an apparent density of approximately 0.17 g/cm$^3$.

\section{Wet gel synthesis}
\label{2}

Wet gel tiles were synthesized by sol--gel chemistry based on the two-step method (e.g., \cite{cite17}). Here, the first step was omitted by purchasing the silica precursor, polymethoxy siloxane [CH$_3$O[Si(OCH$_3$)$_2$O]$_n$CH$_3$, methyl silicate 51; Fuso Chemical Co., Ltd., Japan] \cite{cite8}. The wet gel was simply synthesized via the following successive sol--gel condensation polymerization reaction: CH$_3$O[Si(OCH$_3$)$_2$O]$_n$CH$_3$ + ($n$ + 1)H$_2$O $\to $ (SiO$_2$)$_n$ + (2$n$ + 2)CH$_3$OH. The reaction was base-catalyzed by aqueous ammonia. To obtain highly transparent aerogel tiles, a mixture of methanol and $N$,$N$-dimethylformamide (DMF) was used as the diluent solvent, as suggested in Ref. \cite{cite9}.

To optimize the refractive index and maximize the transparency, the dose of raw chemicals for the wet gel synthesis was updated at each production, based on the result of the previous production. For the mixing ratio in the first production, we referred to our previous study using small size aerogel tiles \cite{cite9}. Some of the latest recipes used in the present study are listed in Table \ref{table:table1}. The amount of basic catalyst was preadjusted to complete the gelation approximately 2 min after the beginning of the polymerization reaction at room temperature. The total volume of the chemicals was predetermined to yield 20 mm-thick aerogel tiles after wet gel shrinkage.

\begin{table*}[t]
\centering 
\caption{Chemical solutions used in the wet gel synthesis (per 100 mL wet gel volume). The chemicals were purchased from Wako Pure Chemical Industries, Ltd., Japan, except for polymethoxy siloxane. The numeric components of the identifiers denote the target refractive index.}
\label{table:table1}
	\begin{tabular}{llllllll}
		\hline
		Chemicals & \multicolumn{7}{l}{Dose [g] per wet gel volume of 100 mL} \\
		\cline{2-8}
		 & 1.045A & 1.045B & 1.045C & 1.050D & 1.055B & 1.055D & 1.055F \\
		\hline
		Polymethoxy siloxane & 20.81 & 20.34 & 20.58 & 22.73 & 24.28 & 25.22 & 24.52 \\
		Distilled water & 22.40 & 21.83 & 21.93 & 22.87 & 23.37 & 23.58 & 23.41 \\
		Methanol & 7.09 & 7.20 & 7.17 & 4.56 & 4.42 & 4.34 & 4.40 \\
		$N$,$N$-Dimethylformamide & 48.32 & 49.11 & 48.87 & 49.39 & 47.85 & 46.98 & 47.64 \\
		28\% Ammonia solution & 0.23 & 0.22 & 0.22 & 0.20 & 0.19 & 0.20 & 0.19 \\
		\hline
	\end{tabular}
\end{table*}

In the first and second productions, commercially-available larger polystyrene molds (type 12--14) were used, with a conventionally sized mold (type 19) as the reference; thereafter (in production runs 3--7), these were replaced with a custom-made polystyrene mold (type A) with inner dimensions of 187 $\times $ 187 mm$^2$. Table \ref{table:table2} lists the specific sizes of each mold. The size of the custom-made mold was determined from the results of the 1st and 2nd productions. To maximize the tile width, the trade-off between the aerogel radiator tiling scheme and the capacity of the supercritical drying apparatus was considered.

\begin{table*}[t]
\centering 
\caption{Polystyrene molds used in the wet gel synthesis.}
\label{table:table2}
	\begin{tabular}{lll}
		\hline
		Identifier & Inner dimensions [mm$^2$] & Supplier \\
		\hline
		Type 14 & 327 $\times $ 235 & As One Corporation, Japan \\
		Type 13 & 291 $\times $ 203 & id. \\
		Type 12 & 264.5 $\times $ 185.0 & id. \\
		Type 19 & 155.5 $\times $ 155.5 & id. \\
		Type A & 187.0 $\times $ 187.0	 & Saito Shoten Ltd., Japan (custom fabrication) \\
		\hline
	\end{tabular}
\end{table*}

Solution A was first prepared by adding methanol and DMF to polymethoxy siloxane in a beaker. Next, water was mixed with aqueous ammonia in another beaker to yield Solution B. Immediately after adding solution B to solution A, the mixed solution was stirred for 20 s, and then poured into the mold. After several minutes, the formed wet gel was covered with a small amount of methanol, and the mold was covered with an aluminum (or polystyrene) lid to prevent the gel from drying. The wet gel in the covered mold was transferred to a sealed tank and aged, typically for two week. After removing the lid, the tank was filled with 2-propanol, and the wet gel in the mold was immersed in the tank for 1--3 h. The wet gel was then detached from the mold, transferred to a stainless steel punched tray, and again immersed in the 2-propanol tank for 1--3 days. In the later processes, the wet gel tile was handled through the punched tray without touching it directly.

In this study, the hydrophobic treatment was performed by the hydrophobic reagent hexamethyldisilazane [((CH$_3$)$_3$Si)$_2$NH; Dow Corning$^{\textregistered }$ Z-6079 Silazane; Dow Corning Toray Co., Ltd., Japan] prior to supercritical drying (SCD). In contrast, the hydrophobic treatment and SCD were simultaneously performed in Ref. \cite{cite18} (see also Ref. \cite{cite5}). The hydrophobic reaction at the surface of the silica clusters is given by: 2(--OH) + ((CH$_3$)$_3$Si)$_2$NH $\to $ 2(--OSi(CH$_3$)$_3$) + NH$_3$. The hydrophobic reagent was poured into a new 2-propanol tank at a hexamethyldisilazane concentration of approximately 10 v/v\% in 2-propanol, and stirred. The wet gel in the punched tray was statically soaked in the hydrophobic solution for 2--3 days, retrieved from the solution and immersed in new 2-propanol. This wash step removed impurities such as ammonia generated in the hydrophobic reaction and unreacted chemicals. The washing process was repeated twice (i.e., a total of three washes) within approximately one week.

In productions 1--4, the wet gel processing at room temperature was conducted by Panasonic Electric Works Co., Ltd., Japan. In the 5th to 7th productions, it was carried out by the Japan Fine Ceramics Center, which inherited the aerogel manufacturing business from Panasonic Electric Works in 2012. The production technique for the wet gel tiles was based on the KEK--Chiba--Panasonic standard procedure (i.e., the modified KEK method) \cite{cite2}, originally developed by the KEK--Matsushita group \cite{cite8} (Matsushita Electric Works Co., Ltd., Japan, the predecessor of Panasonic Electric Works).

\section{Supercritical carbon dioxide drying}
\label{3}

The synthesized wet gel tiles were subjected to supercritical carbon dioxide (scCO$_2$) drying in a dedicated supercritical extraction setup with a 140 L autoclave (Mohri Oil Mill Co., Ltd., Japan). First built in the 1980s, the extraction system using scCO$_2$ was installed at KEK to produce low-refractive-index aerogel radiator blocks for the aerogel Cherenkov counters of the Belle detector. Since its transfer to Mohri Oil Mill, the SCD has been managed by that company. The use of the SCD system was planned for manufacturing the aerogel radiator tiles for the A-RICH counter of the Belle II detector. To store the punched trays containing the wet gel tiles, a cylindrical mesh cage containing racks with approximate diameters of 41 cm fitted with the autoclave was used. Thus, the aerogel radiator tiling design of the A-RICH counter was constrained by the rack dimensions.

The wet gel tiles in the punched trays were placed on the racks in the cage. The cage was stored in a sealed tank filled with new 2-propanol, which was then shipped from the Panasonic Electric Works factory at Iga or the Japan Fine Ceramics Center laboratory at Nagoya to the Mohri Oil Mill factory at Matsuzaka. The shipping vehicle was a suspension truck (the travel distance was approximately 100 km). At Mohri Oil Mill, the cage containing the wet gel tiles was transferred to the autoclave filled with new 2-propanol.

\begin{figure}[t]
\centering 
\includegraphics[width=0.49\textwidth,keepaspectratio]{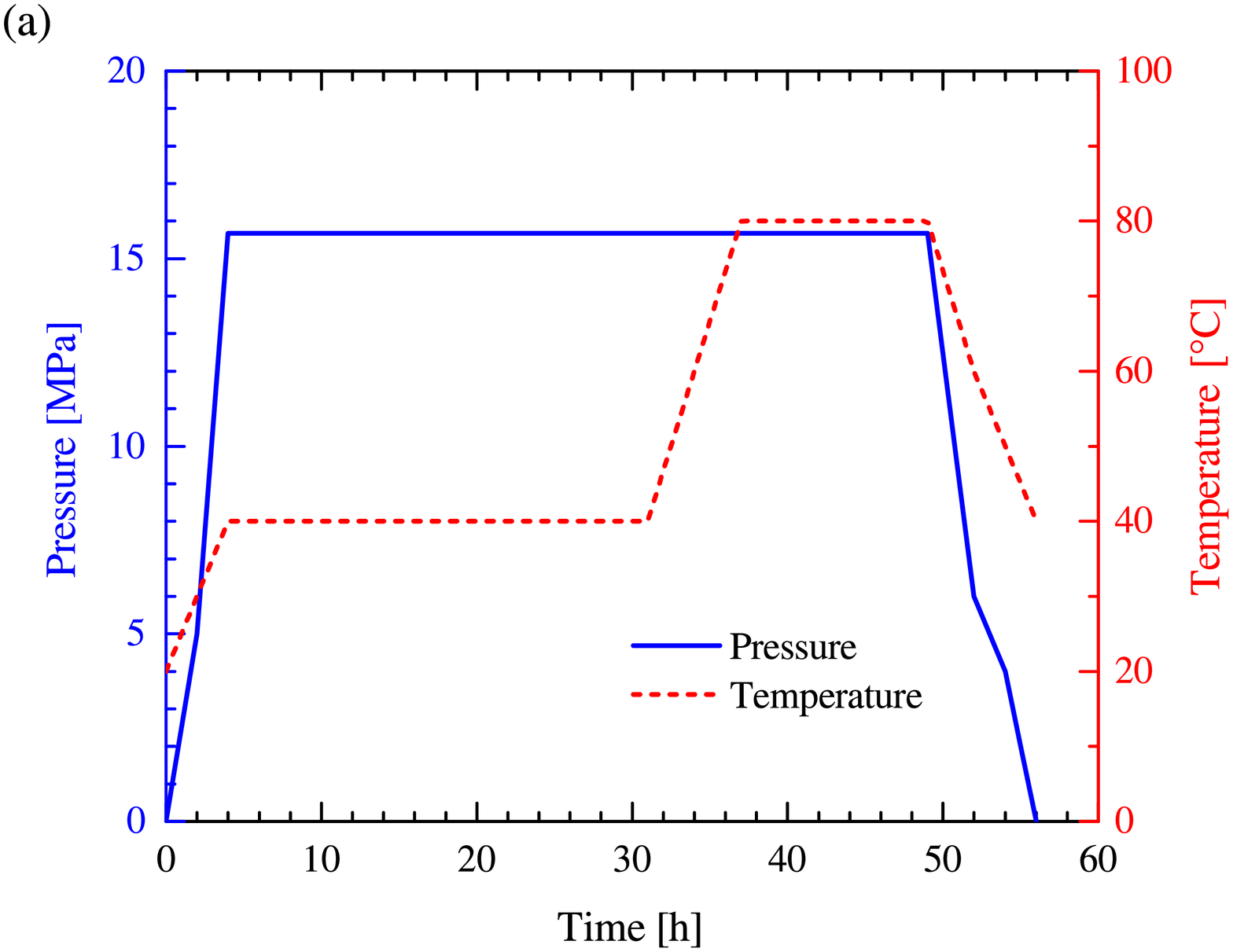}
\includegraphics[width=0.49\textwidth,keepaspectratio]{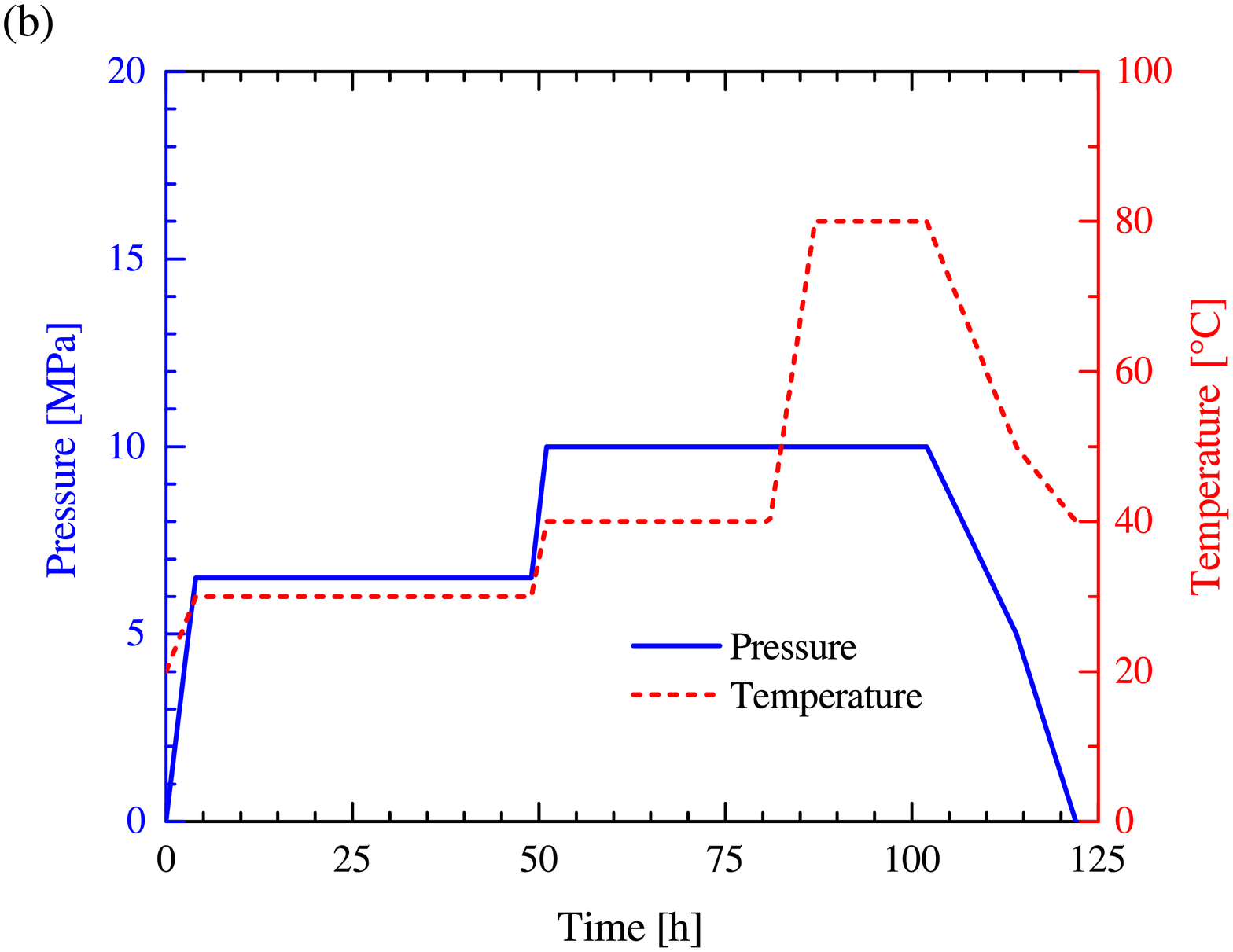}
\caption{SCD design process in (a) the conventional (KEK--Panasonic--Mohri standard) and (b) the modified (Chiba) patterns. The solid and dashed lines indicate the pressure and temperature profiles in the autoclave, respectively.}
\label{fig:fig2}
\end{figure}

The conventional SCD design process (i.e., the KEK--Panasonic--Mohri standard) is shown in Fig. \ref{fig:fig2}a. As is well known, the critical temperature and pressure of CO$_2$ are approximately 31$^{\circ }$C and 7.4 MPa, respectively \cite{cite19}. In the primary pressurization, the autoclave pressure is raised from atmospheric pressure to 5 MPa, and the autoclave temperature is raised from 20 to 30$^{\circ }$C over a period of 2 h. In the secondary pressurization, the pressure is raised from 5 to 15.7 MPa, and the temperature is raised from 30 to 40$^{\circ }$C over a further 2 h period. Maintaining the final conditions (40$^{\circ }$C, 15.7 MPa), the autoclave is pumped with CO$_2$, and the mixed 2-propanol/CO$_2$ fluid is extracted. This process continues for 27 h. Next, the temperature is raised from 40 to 80$^{\circ }$C over a period of 6 h, and the final conditions (80$^{\circ }$C, 15.7 MPa) are maintained while pumping CO$_2$ and extracting the fluid for a further 12 h. After confirming that the 2-propanol concentration in the extracted fluid is below 100 ppm, the CO$_2$ pumping is halted. In the primary depressurization, the pressure is lowered from 15.7 to 6 MPa, and the temperature is lowered from 80 to 60$^{\circ }$C over a period of 3 h. In the secondary depressurization, the pressure is lowered from 6 MPa to atmospheric pressure, and the temperature is lowered from 60 to 40$^{\circ }$C over a period of 4 h. The conventional SCD process requires 56 h to obtain the final aerogel products.

Based on studies undertaken at Chiba University \cite{cite2}, the SCD design process was modified as shown in Fig. \ref{fig:fig2}b. In the primary pressurization, the pressure is raised from atmospheric pressure to 6.5 MPa, and the autoclave temperature is raised from 20 to 30$^{\circ }$C over a period of 4 h. The conditions (30$^{\circ }$C, 6.5 MPa) are statically maintained for 45 h. In the secondary pressurization, the pressure is raised from 6.5 to 10 MPa, and the temperature is raised from 30 to 40$^{\circ }$C over a period of 2 h. The new conditions (40$^{\circ }$C, 10 MPa) are maintained while pumping CO$_2$ and extracting the fluid for 30 h. The temperature is then raised from 40 to 80$^{\circ }$C over a period of 6 h, and the final conditions (80$^{\circ }$C, 10 MPa) are maintained while pumping CO$_2$ and extracting the fluid for 15 h. After checking the 2-propanol concentration, the CO$_2$ pumping is halted, and the primary depressurization is started. Here, the pressure is lowered from 10 to 5 MPa, and the temperature is lowered from 80 to 50$^{\circ }$C over a period of 12 h. In the secondary depressurization, the pressure is lowered from 5 MPa to atmospheric pressure, and the temperature is lowered from 50 to 40$^{\circ }$C over a period of 8 h. The total time of the modified SCD process is 122 h.

\section{Optical characterization methodology}
\label{4}

All of the fabricated aerogel tiles were visually inspected for signs of cracking, surface damage, and abnormal coloration. Their optical parameters (i.e., refractive index and transmittance) were then measured tile by tile. The dimensions and weight of each tile were also measured with a ruler and an electronic balance, respectively. In contrast, the uniformity of the tile density was evaluated in three selected tiles.

\subsection{Refractive index measurement}
\label{4-1}

The apparent refractive index of the aerogel tiles was measured by the minimum deviation method (also referred to as the Fraunhofer method \cite{cite2}) using a laser; specifically, a blue--violet semiconductor laser with $\lambda $ = 405 nm (MLXA-A16-405-5; Kikoh Giken Co., Ltd., Japan). The wavelength of the laser device approximates the wavelength at which typical photodetectors exhibit their peak quantum efficiency. The four corners (regarded as 90$^{\circ }$) of each tile were irradiated with the laser beam, and the minimum laser deviation was measured on a screen approximately 1.8 m downstream of the aerogel tile. Without the aerogel tile, the laser spot on the screen was adjusted to an optimal diameter below 1 mm. When passing through the aerogel tile, the laser beam was diffused, and the increased spot size indicated the optical quality of the tile. Poor optical quality is a major source of measurement error. In addition, the refractive indices might differ at the four corners of individual tiles; therefore, the tile refractive index was defined as the average of the four measured values. In the present study, the refractive index measured in air ($n_{\rm {air}}$) is simply referred to as $n$.

\subsection{Transmittance measurement}
\label{4-2}

\begin{figure}[t]
\centering 
\includegraphics[width=0.49\textwidth,keepaspectratio]{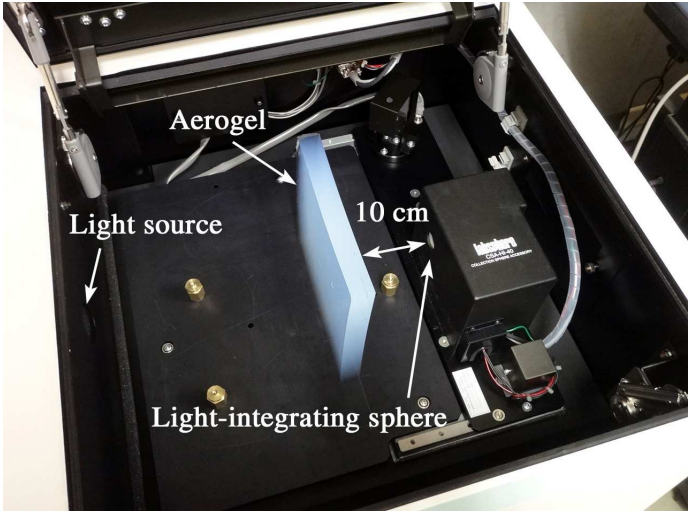}
\caption{Measurement setup in the light-shielded chamber of the Hitachi U-4100 spectrophotometer. The downstream surface of the aerogel was positioned 10 cm from the entrance of the light-integrating sphere.}
\label{fig:fig3}
\end{figure}

The optical transmittance [ultraviolet--visible (UV--VIS) spectrum] of the aerogel tiles was measured at $\lambda $ = 200--800 nm using a spectrophotometer (U-4100; Hitachi, Ltd., Japan). The measurement setup in the light-shielded chamber of the spectrophotometer is shown in Fig. \ref{fig:fig3}. Light transmission was measured along the direction of aerogel thickness. To minimize the capture of light scattered in the aerogel, the bottom surface of the aerogel tile was placed 10 cm from the entrance of the light-integrating sphere (with an entrance diameter of 20 mm) \cite{cite2}. Without the aerogel tile, the spot of the light source spread to approximately 10 $\times $ 8 mm$^2$ at the entrance of the integrating sphere.

\subsection{Density uniformity measurement}
\label{4-3}

The density uniformity of the aerogel tiles was evaluated by a proposed X-ray technique \cite{cite20}. Measurements were conducted in the transverse planar direction (i.e., over the tile face, not the thickness direction). The method exploits the exponential attenuation law of X-ray absorption by materials:
\begin{equation}
\label{eq:eq2}
I/I_0=\exp(-\mu_{\rm m}\rho t),
\end{equation}
where $I/I_0$ is the X-ray transmittance, $\mu_{\rm m}$ is the X-ray mass absorption coefficient, and $t$ is the material thickness. These three parameters were measured, and the aerogel density $\rho $ was obtained by rearranging Eq. (\ref{eq:eq2}). The X-ray transmittances were measured by a NANO-Viewer (Rigaku Corporation, Japan), a measurement system that measures the X-ray absorption. Each tile was subjected to a spot X-ray irradiation measurement; i.e., its density distribution was scanned along its diagonal by a monoenergetic narrow X-ray beam.

The X-ray mass absorption coefficients of the aerogel tiles were determined from the weight fractions and absorption coefficients of their atomic constituents. By calculating the absorption coefficient of the aerogel, the absolute density can be obtained at each X-ray impact point. In the present study, the absorption coefficient of aerogel synthesized in DMF solvent was used, which was previously determined for another typical aerogel sample \cite{cite20}. In the previous study, the atomic fraction of the aerogel was measured in an X-ray fluorescence spectrometer (ZSX100e; Rigaku Corporation).

The aerogel thickness was measured by a universal measuring microscope (UMM200, Tsugami Corporation, Japan). To expose the cross section in the thickness direction along the X-ray impact positions, the aerogel tile was cut along its diagonal with a water jet cutter, allowing precise thickness measurements. The integrity of the aerogel characteristics during water jet machining is discussed in Section \ref{5-4}.

\section{Results and discussion}
\label{5}

\subsection{Cracking issue}
\label{5-1}

The present study includes 7 experimental aerogel productions comprising 13 SCD batches (269 manufactured tiles in total). Table \ref{table:table3} lists the production details and the counted numbers of crack-free aerogel products. In this subsection, the detachment of the wet gel tile from the mold is first described, then, the bubble issue in the wet-gel synthesis process is discussed. Finally, the cracking issue in each production is detailed.

The polystyrene mold material enhanced the detachment of the wet gel tiles from the mold. During the aging period, the wet gel tile slightly shrank and spontaneously separated from the mold wall, facilitating its detachment from the mold. Once separated from the mold, the tile was easily placed into the punched tray by overturning the mold in the 2-propanol bath. This wet-gel detachment procedure was carefully performed to prevent the wet gel from cracking or chipping at the tile corner. The aerogel tiles from the latest batch had no chipping at all because the wet gel tiles were carefully handled. The polystyrene mold was essential to this success.

When mixing solution A with solution B in the wet gel synthesis, the agitation intensity during the 30 s stirring step largely influenced the unintended bubble formation inside the wet gel at the gelation step. Tiny bubbles formed at this stage broadened and flattened in the aging process, causing internal cracking of the gel structure. Thus far, to prevent bubbles in the wet gel, the mixed solution was moderately stirred; however, it was difficult to effectively remove the bubbles. The aerogel products of the first production were seriously cracked because of bubble formation. Conversely, according to an investigation at Chiba University, strong stirring effectively removed the bubbles in the gelation step. Although the agitation intensity was not quantitatively measured, the solution was forcefully stirred until bubbles formed in the beaker. Once the solution containing the bubbles was poured into the mold, the bubbles were spontaneously evacuated from the sol surface no later than the complete gelation. In later productions, the bubbles were well suppressed.

\begin{table*}[t]
\centering 
\caption{Summary of specifications and results of the experimental productions.}
\label{table:table3}
	\begin{tabular}{lllll}
		\hline
		SCD batch & SCD process & Mold type & Crack-free/all products [tile] & Remarks \\
		\hline
		1st (3 batches) & Conventional & Type 14 & 0/10 (0\%) & \\
		 &  & Type 13 & 0/12 (0\%) & \\
		 &  & Type 12 & 2/12 (17\%) & \\
		 &  & Type 19 & 3/6 (50\%) & \\
		2nd (batch 1) & Chiba type & Type 12 & 5/12 (42\%) & \\
		 &  & Type 19 & 5/6 (83\%) & \\
		2nd (batch 2) & Conventional & Type 12 & 2/12 (17\%) & \\
		 &  & Type 19 & 5/6 (83\%) & \\
		2nd (batch 3) & Chiba type & Type 12 & 4/8 (50\%) & \\
		 &  & Type 19 & 7/8 (88\%) & \\
		3rd & Chiba type & Type A & 4/28 (14\%) & Pressure control error \\
		4th & Chiba type & Type A & 14/28 (50\%) & \\
		5th (batch 1) & Chiba type & Type A & 27/28 (96\%) & Chromatic contamination \\
		5th (batch 2) & Chiba type & Type A & 26/28 (93\%) & \\
		6th (batch 1) & Chiba type & Type A & 25/28 (89\%) & \\
		6th (batch 2) & Chiba type & Type A & 6/9 (67\%) & Pressure control error \\
		7th & Chiba type & Type A & 22/28 (79\%) & \\
		\hline
	\end{tabular}
\end{table*}

\subsubsection{First production}
\label{5-1-1}

In the first production, a total of 40 wet gel tiles was dried by the \textit {conventional} SCD process in three sequential batches. The molds were the rectangle types 14, 13, and 12 and the conventional square type 19 as a reference. All of the aerogel tiles obtained from mold types 14 and 13 were cracked, and only 2 of the 12 tiles obtained from mold type 12 were crack-free. These wet gel tiles were difficult to handle without cracking on account of their large area and asymmetric (i.e., rectangular) shape. That is, many of the wet gel tiles had internal cracks before the SCD. In the reference mold, 3 out of 6 tiles were crack-free. We considered that crack-free aerogel tiles would be difficult to produce in mold types 14 and 13, even if the SCD procedure was improved in the next production.

\subsubsection{Second production}
\label{5-1-2}

Based on our experiences with homebuilt supercritical drying apparatus, including the 7.6 L autoclave at Chiba University \cite{cite2}, we modified the SCD process in later productions. We focused on mold type 12, retaining mold type 19 as the reference. The concept of the modified procedure is slow pressure control. First, the time of the primary pressurization step was doubled from 2 to 4 h and increased the final pressure of this step from 5 to 6.5 MPa. The conditions were then statically maintained at 30$^{\circ }$C and 6.5 MPa for 45 h, so that the liquefied CO$_2$ was well mixed with 2-propanol. At Chiba University, this static customization step required approximately 12 h; thus, the additional 33 h may be redundant in the present study. During this step, the moderate solvent exchange from 2-propanol to liquefied CO$_2$ relaxed the stress on the silica matrix. In the secondary pressurization step, the maximum SCD pressure was lowered to 10 MPa (previously to 15.7 MPa). The pumping rate of scCO$_2$ was then slightly decreased. Finally, to suppress cracking, the pressure was very slowly reduced from 10 MPa to atmospheric over a period of 20 h (previously from 15.7 MPa over a period of 7 h).

The modified SCD process showed promise in suppressing the aerogel cracking. In the second production, batches 1 and 3, which were dried by the modified process, were compared with batch 2, which was dried by the conventional process. The wet gel tiles were on the whole crack free before the SCD. In the conventional process using mold type 12 (batch 2), only 2 out of 12 tiles were crack-free, reproducing the result of the first production. The modified process increased the crack-free tile yield to 50\%; i.e., batch 1 (3) yielded 5 (4) crack-free tiles out of 12 (8) tiles. Tiles dried by the conventional and modified processes showed the same optical performances. Hereafter, the modified (Chiba type) SCD process was employed as our standard.

\subsubsection{Productions 3--7}
\label{5-1-3}

Accounting for all possible sizes of the aerogel tiles, the autoclave size, and the cylindrical shape of the end cap of the Belle II detector, the radiator tiling plan presented in Fig. \ref{fig:fig1} was devised. This plan requires 248 aerogel tiles (plus spares) with dimensions of 18 $\times $ 18 $\times $ 2 cm$^3$. The SCD apparatus could dry only 12 type-12 tiles (i.e., 1 tile × 12 columns) per batch. We manufactured a custom-made polystyrene mold (type A) with inner dimensions of 187 $\times $ 187 $\times $ 30 mm$^3$ and a dedicated stainless steel punched tray with inner dimensions of 197 $\times $ 197 $\times $ 24 mm$^3$, which just fitted the type-A wet gel tile. In this apparatus, a nominal 28 wet gel tiles (i.e., 2 tiles $\times $ 14 columns) per batch could be dried. The length of mold type A approximately equaled the shorter length of mold type 12. In productions 3--7, only mold type A was used.

\begin{figure}[t]
\centering 
\includegraphics[width=0.49\textwidth,keepaspectratio]{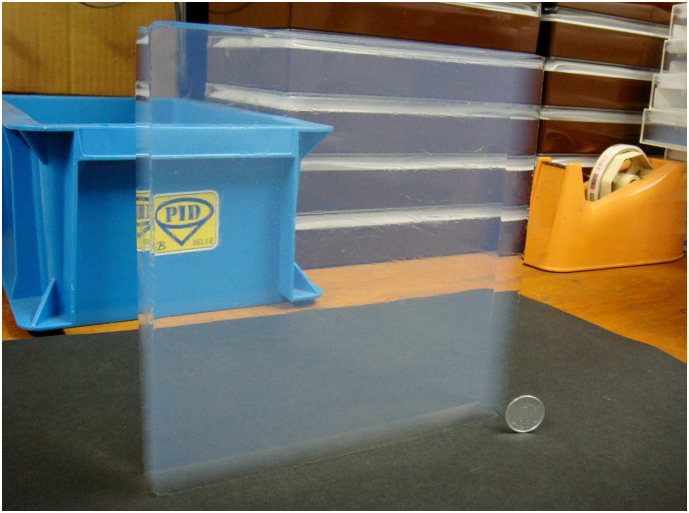}
\caption{The first mold-type-A aerogel sample with $n$ = 1.0446 and dimensions of 18 $\times $ 18 $\times $ 2 cm$^3$. The coin (diameter = 2 cm) is displayed as a reference.}
\label{fig:fig4}
\end{figure}

The third production confirmed that slow pressure reduction at the end of the SCD process improved the crack-free tile yield. In the primary depressurization, the pressure must be lowered from 10 to 5 MPa over a period of 12 h; however, the pressure in the third production was reduced in only 5 h due to improper operation of the automated output valve in the autoclave. Consequently, only 4 out of 28 tiles (14\%) were crack-free. However, the first successful tile synthesized in mold type A was attained, as shown in Fig. \ref{fig:fig4}. Since then, the output valve has been maintained to ensure its correct operation.

The fourth production achieved 14 crack-free tiles out of 28 tiles (50\% yield). Among the 6 batches in productions 4--7, the crack-free yield was maximized at 96\%. In the 5th production, the pressurization was moderated by automating the CO$_2$ input valve to the autoclave, improving the crack-free yield to 80\% or higher. The automatic valve consistently achieved a high crack-free yield (see Table \ref{table:table3}). Besides the slow depressurization, the moderate pressurization using the automated input valve played an important role in suppressing aerogel cracking, as revealed in batch 2 of the sixth production. In this batch, the recorded pressure profile deviated from the planned profile due to an accidental operation error of the input/output valves; consequently, the crack-free yield was relatively low (6 out of 9 tiles; 67\%). By properly operating the SCD apparatus, a method for mass-producing large-area, crack-free aerogel tiles was thereafter established.

\subsubsection{Cracking types}
\label{5-1-4}

\begin{figure}[t]
\centering 
\includegraphics[width=0.49\textwidth,keepaspectratio]{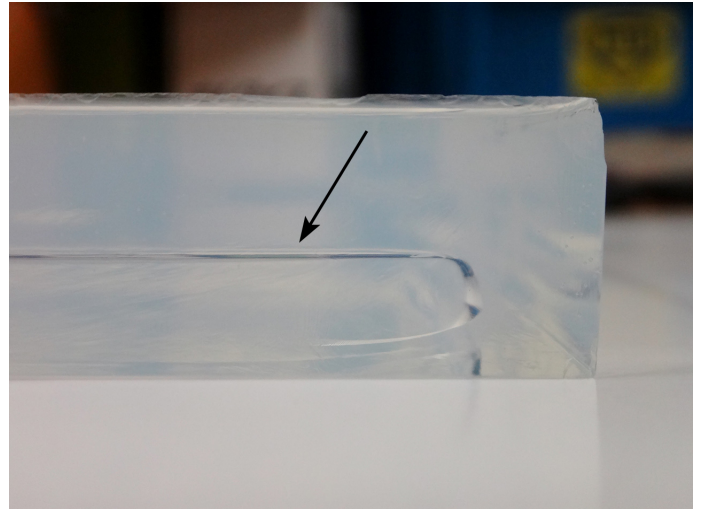}
\caption{Transverse crack generated along the half-height plane of the aerogel thickness. The cracking plane is indicated by the arrow in the aerogel side view.}
\label{fig:fig5}
\end{figure}

There were two types of cracks in the manufactured aerogel tiles: random and transverse. Aerogel tiles could develop either or both of these crack types. Random cracks are caused by mechanical damage to the wet gel tiles, either by careless wet gel handling and/or by the SCD process. Transverse cracks appear along the transversal plane approximately halfway through the aerogel thickness, as shown in Fig. \ref{fig:fig5}. Transverse cracks appeared only after the SCD, indicating that they were introduced by the supercritical fluid extraction. The detailed mechanism of transverse cracking remains unsolved, but it may be related to the density profile along the thickness direction. Overall, cracking was well suppressed by the modified SCD process.

\subsection{Refractive index and transmission length}
\label{5-2}

\subsubsection{Basic optical performance}
\label{5-2-1}

\begin{figure}[t]
\centering 
\includegraphics[width=0.49\textwidth,keepaspectratio]{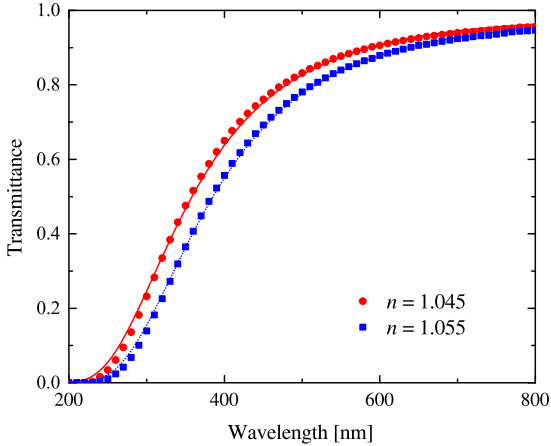}
\caption{UV--VIS spectra of typical 20 mm-thick aerogel tiles with $n$ = 1.045 (circles) and 1.055 (squares). The solid and dotted curves are the fits to $n$ = 1.045 and 1.055, respectively, using Eq. (\ref{eq:eq3}).}
\label{fig:fig6}
\end{figure}

The clarity parameters of the aerogel tiles were determined from UV--VIS measurements. The optical transmittance ($T$) of aerogel is known to be dominated by Rayleigh scattering, expressed as:
\begin{equation}
\label{eq:eq3}
T(\lambda ,t)=A\exp (-Ct/\lambda ^4),
\end{equation}
where $A$ is the amplitude and $C$ is called the \textit {clarity coefficient}, usually measured in units of $\mu $m$^4$/cm. To a minor extent, light transmission in the aerogel is also affected by absorption and surface scattering. Figure \ref{fig:fig6} shows a plot of the transmittance curves of typical aerogel tiles with $n$ = 1.0450 ($t$ = 20.1 mm) and 1.0550 (20.3 mm) which were obtained in the fifth production (batch 2; mold type A) as a function of wavelength measured every 10 nm. By fitting these data to Eq. (\ref{eq:eq3}), $A$ = 0.990 $\pm $ 0.002 (0.987 $\pm $ 0.002) and $C$ = 0.00566 $\pm $ 0.00005 (0.00755 $\pm $ 0.00004) $\mu $m$^4$/cm were obtained for the tiles with $n$ = 1.0450 (1.0550).

\begin{figure}[t]
\centering 
\includegraphics[width=0.49\textwidth,keepaspectratio]{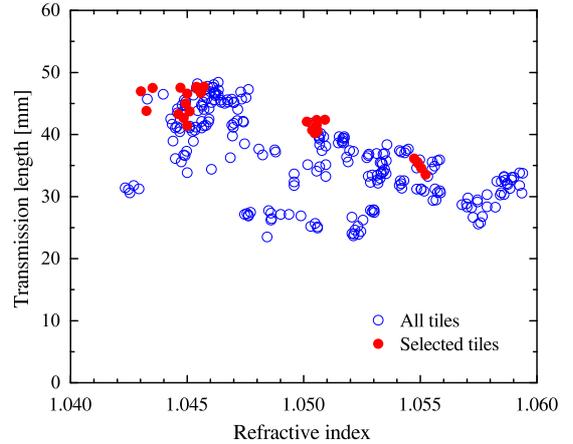}
\caption{Transmission lengths calculated at $\lambda $ = 400 nm as a function of the refractive index. The apparent refractive index was measured at $\lambda $ = 405 nm using a laser. Circles are the measured data of all 266 aerogel tiles. Among these, the selected tiles (28 samples) are highlighted by closed circles.}
\label{fig:fig7}
\end{figure}

Another parameter that determines the optical transparency of the aerogel tiles is the transmission length $\Lambda _T$, defined as $\Lambda _T(\lambda ) = -t/\ln T(\lambda )$. The $\Lambda _T$ was calculated from the UV--VIS data at $\lambda $ = 400 nm and is plotted as a function of the apparent refractive index in Fig. \ref{fig:fig7}. The refractive index was measured by directing a 405-nm laser at the tile corners. The data of all measured aerogel tiles (266 samples) are plotted, and the 28 selected tiles are highlighted by closed circles. The selected aerogel samples were assessed as good (i.e., suitable for the A-RICH counter) tiles synthesized by the fully developed sol--gel recipes (identifiers 1.045B, 1.050D, and 1.055D in Table \ref{table:table1}) in mold type A. Tiles with cracks and chromatic contamination (described below), synthesized by the developing sol--gel recipes in molds other than type A, were excluded.

The refractive indices and transmission lengths of the selected tiles satisfied the radiator requirements. Based on an investigation of small-size aerogel tiles \cite{cite2,cite10}, the target transmission lengths in the present study were 45, 40, and 35 mm for $n$ = 1.045, 1.050, and 1.055, respectively. Moreover, the refractive index must be distributed within $\pm $0.002 of the target refractive index. Adjusting the refractive index by investigating the chemical recipes was successful. The refractive indices of the selected tiles targeting $n$ = 1.045 were relatively scattered, as discussed in Section \ref{5-2-2}, but within the scope of the requirements. The transmission lengths of each target refractive index also roughly met our requirements.

In batch 1 of the fifth production, the dried aerogel tiles developed chromatic contamination. Yellow coloration of the tiles after the supercritical drying was clearly visible, and the UV--VIS spectrum noticeably dropped in the $\lambda $ = 200--550 nm range, reducing the transmission lengths. Our 140 L apparatus reused the CO$_2$ used in another supercritical extraction apparatus, to which it was connected, and which introduced the low transmission lengths that were also attained in the fourth production and batch 1 of the sixth production. Although no apparent coloration was observed in these tiles, their amplitudes $A$ (or transmittances at $\lambda $ = 800 nm) were obviously low (typically decreased by 5 percentage points) relative to the healthy tiles), indicating the presence of chromatic contamination. In the seventh production, the problem was resolved by introducing an activated carbon filter to remove contaminants from the CO$_2$.

\subsubsection{Dependence of refractive index on room temperature}
\label{5-2-2}

The apparent refractive index of the manufactured aerogel tiles seems to slightly depend on the temperature in the laboratory of the wet gel synthesis and processing. Although the laboratory temperature is conditioned 24 hours a day throughout the year, it is affected by extreme outside temperatures. In Japan, the outdoor temperature exceeds 30$^{\circ }$C in summer and falls below 5$^{\circ }$C in winter. We now know that temperature especially affects the wet gel shrinkage in the aging process; however, the aging temperature was not precisely controlled in the present study. The selected aerogel tiles in Fig. \ref{fig:fig7} include samples from productions 3--7. Among these, the tiles targeting $n$ = 1.045 include samples from production 4 in summer and production 7 in winter. The mean and standard deviation of the refractive indices of 3 tiles manufactured in summer were 1.0433 and 0.0002 respectively, whereas those of 5 tiles manufactured in winter were 1.0455 and 0.0001, respectively. The refractive index differed most widely between the summer and winter productions. In contrast, the refractive indices of tiles within a batch which were synthesized by the same recipe showed little variation (small standard deviation). 

The difference in the refractive index of aerogel tiles produced in summer and winter reflected the difference in their tile densities, which can be partly explained by differences in the longitudinal tile shrinkage ratio. The mean side lengths of tiles manufactured in summer and winter were 182.5 and 181.5 mm respectively, corresponding to a longitudinal shrinkage ratio of 97.6 and 97.1\% from the mold size of 187.0 mm (type A), respectively. The tile shrinkage (i.e., density increase) contributes half of the difference in the refractive index. The mean densities of samples produced in summer and winter were 0.153 and 0.159 g/cm$^3$, respectively. Considering the relationship between the refractive index and density [i.e., Eq. (\ref{eq:eq1})] and using the parameters given in Ref. \cite{cite2}, the difference in the refractive index is consistent with the density difference, allowing for conversion error.

Apart from tile shrinkage, the nanoscopic structures (silica clusters and pores) of tiles produced in summer or winter could affect the refractive indices. The gelation time, and hence the formation of the nanoscopic gel structure, depends on the room temperature of the wet gel synthesis. Supporting this inference, the aerogel transparency depends on the gelation time; the shorter the gelation time, the more transparent the aerogel. Differences in nanoscopic structure should directly alter the refractive index. Moreover, the increased aerogel density may be attributed to increased numbers of hydrophobic groups added in the hydrophobic treatment, which also depends on the nanoscopic structure.

In the present study, the differences in refractive indices were apparently unaffected by the SCD process. In fact, the 9 tiles dried in batch 2 of the sixth production (in summer) were synthesized as spare tiles during the fifth production (in winter). That is, prior to SCD, the spare wet gel tiles were preserved in the 2-propanol bath for approximately half a year. The mean refractive index of 2 samples synthesized in winter and dried in summer was 1.0505 (the target $n$ of both samples was 1.050). The mean refractive index (and standard deviation) of 5 tiles targeting $n$ = 1.050 from batch 2 of the fifth (winter) production was 1.0504 $\pm $ 0.0002, whereas that of 5 tiles from batch 1 of the sixth (summer) production was 1.0476 $\pm $ 0.0001. Thus, the refractive indices of tiles synthesized in winter and dried in summer matched those of tiles produced only in winter. Samples targeting the other refractive indices yielded similar results, implying that the refractive index was essentially determined before the SCD.

\subsection{Tile density uniformity}
\label{5-3}

\begin{figure}[t]
\centering 
\includegraphics[width=0.49\textwidth,keepaspectratio]{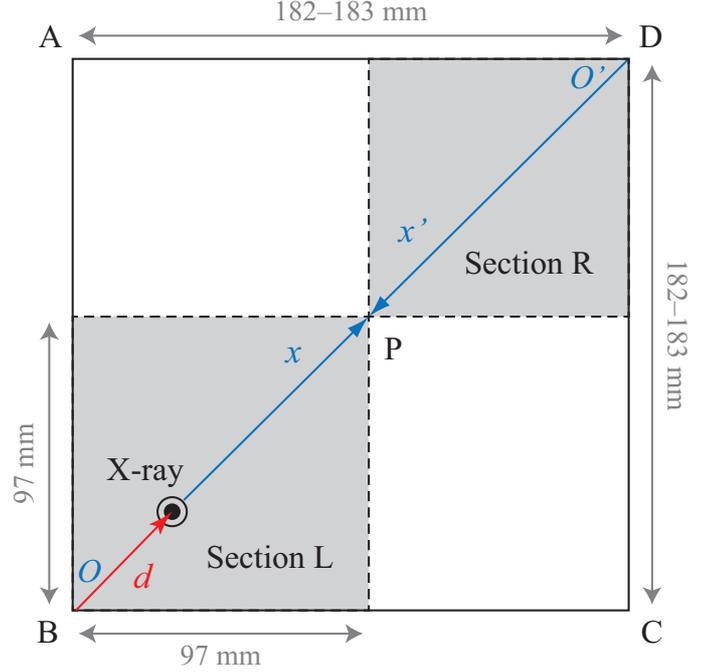}
\caption{Geometric coordinates of the aerogel tile surface synthesized in mold type A. The vertices $A$, $B$, $C$, and $D$ are defined in the counterclockwise direction on the aerogel tile. Dashed lines indicate the tile parting lines, where the tile was cut by a water jet cutter. The two cutting lines intersect at $P$. The areas labeled L and R were subjected to X-ray scan measurements along the $x$- ($x'$-) axis, which connects vertex $B$ ($D$) to $P$. The vertices $B$ (= $O$) and $D$ (= $O'$) of sections L and R respectively were placed at the origin of the aerogel holder, which marks the start point of the X-ray scans. The symbol $d$ defines the distance from $B$ to the X-ray impact position along the $x$- and $x'$-axes.}
\label{fig:fig8}
\end{figure}

The tile density distribution was evaluated in three aerogel samples from the fourth production, synthesized in mold type A. The target (measured) refractive indices of the tiles were 1.045 (1.0433), 1.050 (1.0481), and 1.055 (1.0533). These tiles were cut into four parts along the dashed lines in Fig. \ref{fig:fig8}, and the parts labeled L and R were subjected to X-ray absorption measurements. Cutting was necessary because of limited installation space on the X-ray measurement system; i.e., the whole tile area cannot be scanned at once. Defining vertices $A$, $B$, $C$, and $D$ in the counterclockwise direction on the face of the aerogel tile, the vertices $B$ and $D$ of the L and R sections respectively were placed at the origin of the aerogel holder (denoted $O$ and $O'$, respectively). The origins $O$ and $O'$ coincide on the aerogel holder. To evaluate the transverse planar distribution of the aerogel tile density, X-ray scans were performed along the diagonal line (along the $x$- and $x'$-axes of sections L and R, respectively). The symbol $d$ defines the distance from vertex $B$ to the X-ray impact position.

\begin{figure}[t]
\centering 
\includegraphics[width=0.49\textwidth,keepaspectratio]{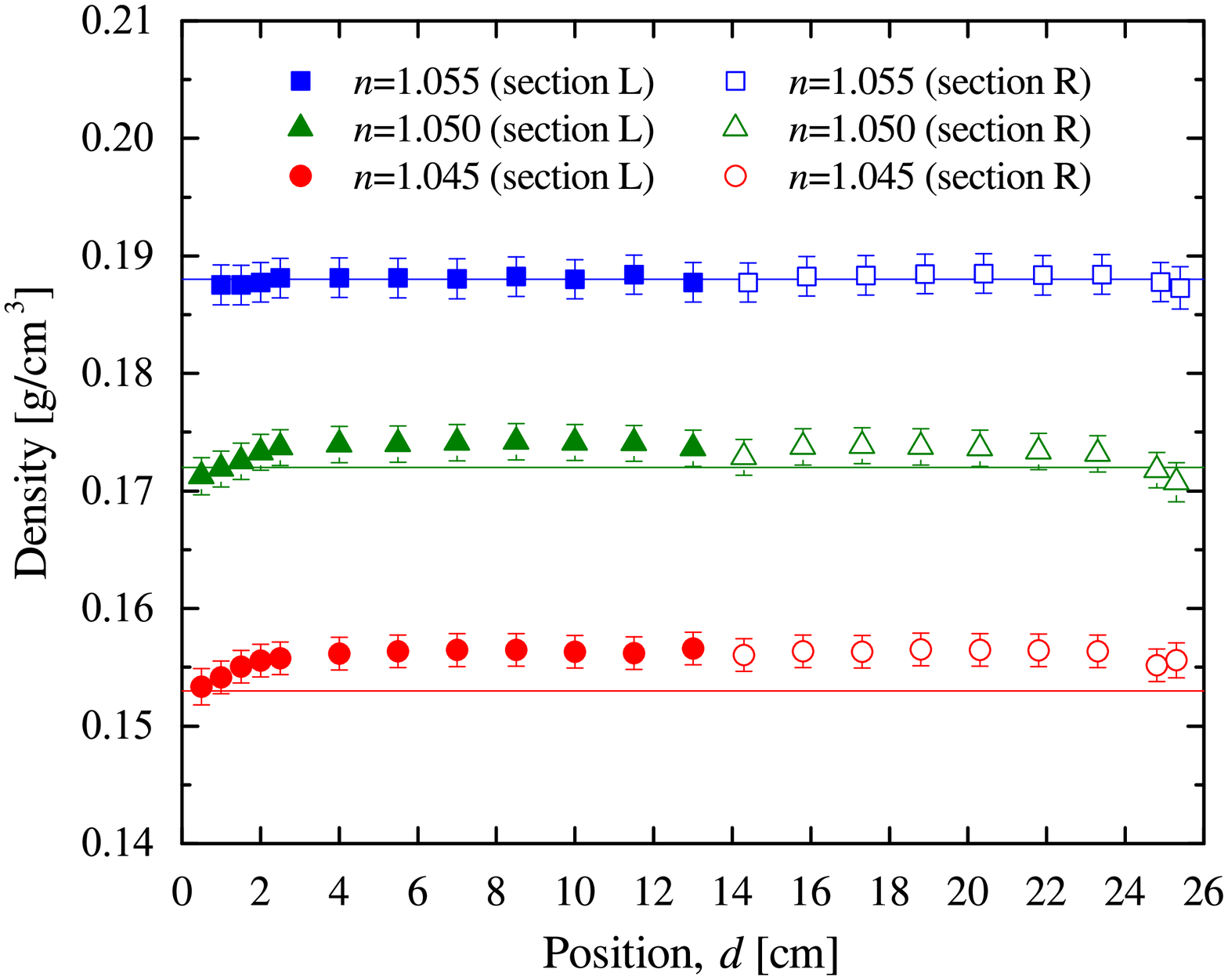}
\caption{Aerogel density versus the X-ray beam position $d$. Circles, triangles, and squares denote aerogel tiles with target refractive indices of 1.045, 1.050, and 1.055, respectively. Closed and open symbols are the data from sections L and R, respectively. Horizontal lines indicate the gravimetric densities of tiles with target refractive indices of 1.045 (bottom), 1.050 (center), and 1.055 (top).}
\label{fig:fig9}
\end{figure}

The density distribution in the diagonal planar direction of each tile was sufficiently flat. Figure \ref{fig:fig9} plots the density as a function of the position $d$, where $d$ = 0 and $d$ = 26 cm approximately correspond to the tile vertices $B$ and $D$, respectively. The horizontal lines near the data of the three tiles indicate the gravimetric densities of the corresponding whole tile. The density measured by the X-ray technique was roughly consistent with the gravimetric density. For the tiles with target refractive indices of 1.045, 1.050, and 1.055, the mean densities around the tile center (10 $\leqq $ $d$ $\leqq $ 16 cm; 5 data) were 0.1563, 0.1737, and 0.1880 g/cm$^3$, respectively. Over most of the tile area (2 $< d <$ 24 cm; excluding the tile corners), the density deviated from the mean by 0.5\% at most. In contrast, at the tile edges (0 $\leqq $ $d$ $\leqq $ 2 cm and 24 $\leqq $ $d$ $\leqq $ 26 cm), the density was up to 2\% lower than the mean. This trend, which approximately mirrored the results of the laser measurements (see Fig. \ref{fig:fig11}), was particularly noticeable in tiles targeting $n$ = 1.045 and 1.050. Nevertheless, the density distributions satisfied our requirements; namely, that $|\delta \rho /\rho| <$ 4\% (i.e., $|\delta (n - 1)/(n - 1)| <$ 4\%). The X-ray measurements are supported by the refractive indices measured at the four corners by the Fraunhofer method; that is, the refractive index measured at a corner was nearly identical to those at the other corners. Similarly, the length of the respective four sides was equivalent each other; thus, the integrity of the tile shape (i.e., square) was nearly confirmed.

\subsection{Hydrophobic characteristics}
\label{5-4}

\begin{figure}[t]
\centering 
\includegraphics[width=0.49\textwidth,keepaspectratio]{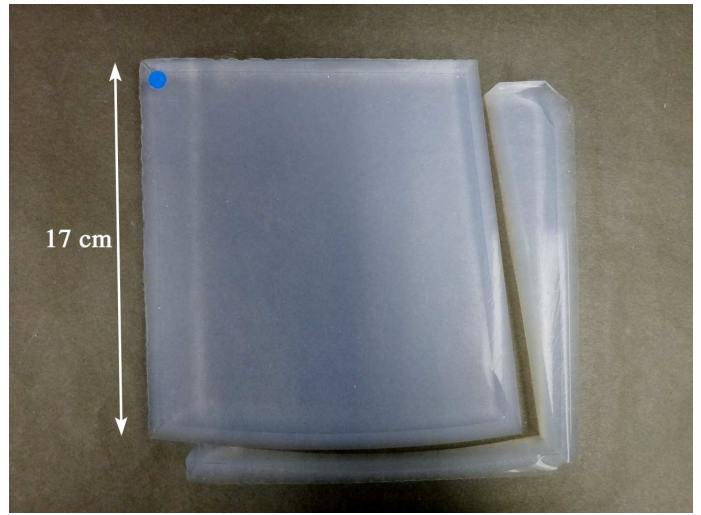}
\caption{Water-jet-machined, partially fan-shaped aerogel tile alongside its separated segment. The sample was cut from an 18 $\times $ 18 $\times $ 2 cm$^3$ tile (synthesized in mold type A) with a target refractive index of 1.055. The top-left corner (indicated by the circular mark) was left untrimmed for measuring the refractive index after the machining. When fully trimmed, this tile will fit the first concentric layer of the end cap in the Belle II detector.}
\label{fig:fig10}
\end{figure}

The hydrophobic characteristics of the manufactured aerogel tiles were confirmed in a water jet machining test. Six tiles (three each targeting refractive indices of 1.045 and 1.055) were selected from productions 5 (batch 2) and 6 (batch 2) and experimentally trimmed by a water jet cutter (Tatsumi Kakou Co., Ltd., Japan). Water jet machining precisely cuts the aerogel tile while exploiting its hydrophobic feature. The aerogel tiles were cut, either fully or partially, into the fan shape that fits the first concentric layer (the layers are counted from the inside of the end cap; see Fig. \ref{fig:fig1}). One of the partially-trimmed tiles is photographed in Fig. \ref{fig:fig10}. Note that a native corner is preserved for measuring the refractive index. The water-jet-machined surface strongly scatters the laser beam used in the refractive index measurement. Therefore, once their edges have been trimmed, the aerogel tiles are unfit for refractive index measurements. To compare the refractive indices before and after machining, partially-trimmed samples targeting each refractive index were thus prepared.

\begin{figure}[t]
\centering 
\includegraphics[width=0.49\textwidth,keepaspectratio]{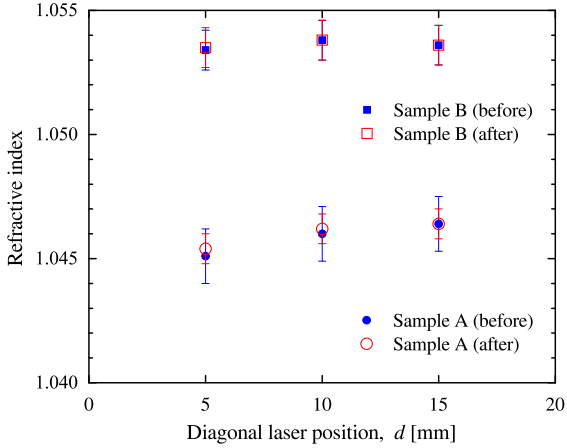}
\caption{Refractive index measured under a 405-nm laser at three diagonal laser positions $d$, where $d$ is defined in Fig. \ref{fig:fig12}. Circles and squares denote aerogel samples A (target $n$ = 1.045) and B (target $n$ = 1.055), respectively. The refractive indices of the designated corner before and after water jet machining are compared by the closed and open symbols, respectively.}
\label{fig:fig11}
\end{figure}

\begin{figure}[t]
\centering 
\includegraphics[width=0.30\textwidth,keepaspectratio]{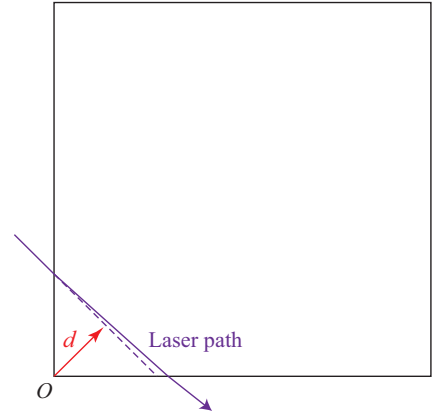}
\caption{Definition of the diagonal laser position $d$. The laser beam (purple solid arrow) cuts across one corner of the aerogel tile. Dashed line indicates the laser path without the aerogel tile. Position $d$ (red solid arrow) is defined as the minimum distance between the aerogel vertex $O$ and the dashed line.}
\label{fig:fig12}
\end{figure}

The refractive index of the aerogel tiles was unaffected by the water jet machining. Figure \ref{fig:fig11} compares the refractive indices of two aerogel samples with $n \sim $1.045 and 1.055 before and after machining. The results are plotted as functions of the diagonal laser position $d$. As illustrated in Fig. \ref{fig:fig12}, $d$ defines the minimum distance between vertex $O$ of the aerogel and the laser path without the aerogel tile, determined by the Fraunhofer method. Using the 405 nm-laser, the refractive index of each aerogel tile was measured at the untrimmed corner of the tile. For full-size tiles synthesized in mold type A, the apparent refractive index was averaged from the refractive indices of the four corners, usually measured at $d$ = 10 mm. In contrast, the refractive index of the partially trimmed tiles could be measured only at one corner. Consequently, for a proper determination, the refractive index at the designated corner was measured at three laser positions ($d$ = 5, 10, and 15 mm). Within the measurement error, the refractive indices before and after the cutting were always consistent, indicating that the aerogel density was not altered by water absorption. This investigation confirmed the hydrophobicity of the aerogel tiles.

\begin{figure}[t]
\centering 
\includegraphics[width=0.49\textwidth,keepaspectratio]{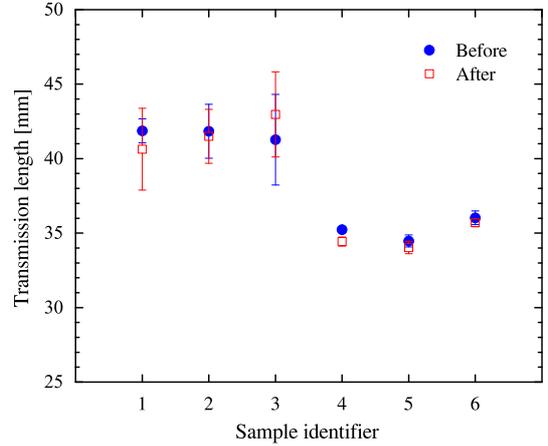}
\caption{Transmission lengths of six aerogel samples, measured at $\lambda $ = 400 nm. Circles and squares indicate the transmission lengths before and after water jet machining, respectively. The error is the standard deviation of the measurements at three tile positions.}
\label{fig:fig13}
\end{figure}

Moreover, the integrity of the optical transmittance of the aerogel samples was preserved after the machining. Figure \ref{fig:fig13} compares the transmission lengths of the 6 aerogel samples at $\lambda $ = 400 nm before and after machining. The six samples are identified by numbers (i.e., 1--6). The transmittance of each tile was measured at three positions on the tile surface (i.e., at near the tile center and left and right sides) and averaged. At these positions, the transmittances were very similar before and after the cutting, but were not identical because the tile shape affects the setup of the aerogel samples in the spectrophotometer. The water jet machining exerted no significant effect on the transmission lengths of the aerogel tiles.

Two characteristic lengths of four fully trimmed aerogel tiles were measured with a ruler. The trimming error in the tile dimensions was below 1\%. Although the true size was slightly smaller than the design size, the dimensions were acceptable and guaranteed safe installation of the aerogel tiles into the radiator support structures. Based on the present study, a second machining test was performed on aerogel tiles that were mass-produced for the actual A-RICH counter. These tiles were successfully installed in a partial mock-up of the aerogel support structure \cite{cite21}.

\section{Conclusion}
\label{6}

Prototypes of silica aerogel tiles have been developed for use in the A-RICH counter of the Belle II detector. The tiles are to be installed as Cherenkov radiators. The A-RICH counter demands large-area aerogel segments to fill the 3.5-m$^2$ cylindrical end cap with the minimum number of two-layer radiators. Based on the results of the present study, the installation of 248 fan-shaped radiator segments trimmed from 18 $\times $ 18 $\times $ 2 cm$^3$ aerogel tiles was planned. Herein, the methodologies for synthesizing the wet gel tiles and rendering them hydrophobic were described. For safe radiator installation and high detector performance, large-area, high-density aerogel tiles with no cracking after the supercritical CO$_2$ drying step are required. The aerogel cracking was effectively suppressed by applying moderate pressure control in the drying procedure. By perfecting our technique, large-area aerogel tiles with a consistently high ($\sim $80\%) crack-free yield were eventually obtained. Moreover, the refractive indices of the aerogel tiles (nominally, $n$ = 1.045, 1.050, and 1.055) were precisely controlled while maintaining high transmission lengths (45, 40, and 35 mm, respectively). In monolithic aerogel tiles with different refractive index targets, the density (i.e., refractive index) distribution along the transverse planar direction was almost uniform, meeting our requirements. The optical parameters were not degraded by water jet trimming, confirming the integrity of their hydrophobic characteristics. Therefore, the aerogel radiator may be feasibly mass-produced for installation in the actual A-RICH counter.

\section*{Acknowledgments}
\label{}
The authors are grateful to Panasonic Corporation, Japan Fine Ceramics Center, and Mohri Oil Mill Co., Ltd. for their contributions to large aerogel production. This paper is the result of many fertile discussions between the authors and Dr. H. Yokogawa of Panasonic Corporation. Prof. K. Nishikawa of Chiba University is particularly to be thanked for her help in X-ray measurements. We are also grateful to Tatsumi Kakou Co., Ltd. for their contributions to water jet machining of aerogel tiles. Thickness measurements of aerogel tiles with the measuring microscope and water jet cutter was performed at the Mechanical Engineering Center at KEK; we are thankful to staff members for their support. Finally, we thank the members of the Belle II A-RICH group for fruitful discussions on aerogel development.

This study was partially supported by a Grant-in-Aid for Scientific Research (A) (No.24244035), a Grant-in-Aid for Scientific Research (C) (No. 19540317), and a Grant-in-Aid for JSPS Fellows (No. 07J02691 for M. Tabata) from the Japan Society for the Promotion of Science (JSPS) and a Grant-in-Aid for Scientific Research on Innovative Areas (No. 21105005) from the Ministry of Education, Culture, Sports, Science and Technology (MEXT). M. Tabata was supported in part by the Space Plasma Laboratory at the Institute of Space and Astronautical Science (ISAS), Japan Aerospace Exploration Agency (JAXA).

\appendix

\section{Requirements of the aerogel Cherenkov radiator in the A-RICH counter}
\label{app}

\subsection{Cherenkov counters}
\label{s1}

The A-RICH counter identifies $B$-meson decay products (e.g., charged $\pi $ and $K$ mesons) by the Cherenkov imaging technique. A charged particle radiates light if its velocity ($v = \beta c$) exceeds the local phase velocity of light (Cherenkov radiation); specifically, if $v > c/n$, where $c$ is the speed of light in a vacuum. The angle $\theta_{\rm {C}}$ of the Cherenkov radiation relative to the particle's direction is given by $\cos \theta_{\rm {C}} = 1/n\beta $. Under the specified refractive index conditions, the A-RICH counter separates kaons from pions using their Cherenkov angle information (i.e., differences in their Cherenkov ring radii) and the supplementary particle velocities (momenta) provided by a charged particle trajectory tracking device \cite{citeS1} in a 1.5-T magnetic field. The A-RICH counter comprises the aerogel Cherenkov radiator tiles, a position-sensitive hybrid photodetector array \cite{citeS2}, dedicated front-end readout electronics \cite{citeS3}, and their mechanical support structures. The Cherenkov ring image that includes the Cherenkov angle information is observed by pixelated photodetectors placed 20 cm downstream of the aerogel radiator surface, and is reconstructed by analyzing the Cherenkov ring radii. The Cherenkov ring images are compiled from the single photon counts obtained by the position-sensitive photodetectors, which are designed to operate in the magnetic field.

In the previous Belle experiment, particle identification at the end cap was achieved by threshold-type aerogel Cherenkov counter (ACC) modules with $n$ = 1.03 radiators \cite{citeS4}. In a single module of the counters with functionality in the magnetic field, the photodetector (comprising two fine mesh-type photomultiplier tubes) was directly attached to the counter-housing filled with aerogel blocks. The ACC separated pions from kaons in the momentum range 0.6--2 GeV/$c$; that is, the $n$ = 1.03 aerogel blocks emitted Cherenkov light only when pions within the specified momentum range passed through the counter module. Particle identification at high momenta (up to 4 GeV/$c$) remained an open problem in the Belle experiment, to be resolved by replacing the ACC system with the A-RICH counter system. To overcome the limited installation space for the A-RICH counter, the A-RICH system was designed as a proximity focusing RICH counter with limited radiator thickness. This design obtains thin Cherenkov ring images without mirror focusing (i.e., introduces an expansion space between the radiator and photodetector) \cite{citeS5}.

\subsection{Optical requirements}
\label{s2}

The refractive index of the aerogel Cherenkov radiators in the A-RICH counter was selected as $\sim $1.05. To observe the Cherenkov light emission triggered by high-energy mesons (charged pions and kaons with momenta up to 4 GeV/$c$) passing through the aerogel, the radiator's refractive index must exceed $\sim $1.008. High $n$ is desirable for increasing the photon emission by the Cherenkov radiator. Conversely, increasing the $n$ generally decreases the transparency of the aerogel \cite{cite2}, degrading the Cherenkov photon collection. As described in the current paper, the upstream surface of the aerogel tiles was separated from the photodetector surface by 20 cm. In radiators with $n$ = 1.05, the Cherenkov angles of 4 GeV/$c$ pions and kaons differ by approximately 23 mrad, corresponding to 5 mm on the photodetection plane. Given the pixel size of the photodetector under development (approximately 5 $\times $ 5 mm$^2$), this difference is detectable. In the 1990s, we achieved the most transparent aerogels to date with $n$ = 1.02--1.03 \cite{cite10,cite11}. In recent studies, we successfully exceeded this transparency with higher $n$ range ($n$ = 1.04--1.05) by applying DMF as the diluent solvent in the sol--gel polymerization step (i.e., by modifying the KEK method) \cite{cite2,cite9,cite10}. Experimentally fabricated small-size aerogel tiles with $n$ = 1.04--1.06 showed promising performance in test beam experiments. However, the photodetector in these tests was a position-sensitive multi-anode photomultiplier tube array, which does not operate in the magnetic field \cite{citeS6,citeS7}.

\begin{figure}[t]
\centering 
\includegraphics[width=0.30\textwidth,keepaspectratio]{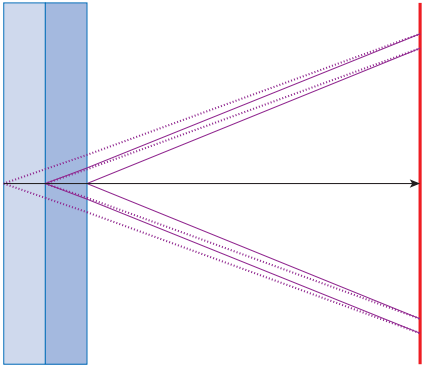}
\caption{Concept of the focusing dual-layer scheme, employing aerogel radiators of different refractive indices. The rectangles at the left and the vertical bold line at the right indicate the dual-layered aerogel tiles and the photodetection plane, respectively. The horizontal arrow shows the charged particle path. Dotted and solid lines denote the Cherenkov lights emitted from the upstream and downstream aerogel layers, respectively. The higher-refractive-index aerogel tiles are arranged behind those with lower refractive index, so that the Cherenkov lights emitted from both layers focus on the photodetection plane.}
\label{fig:figS1}
\end{figure}

The A-RICH counter will employ a focusing dual-layer aerogel radiator scheme with different refractive indices \cite{citeS8}, as illustrated in Fig. \ref{fig:figS1}. The higher-refractive-index aerogel tiles are arranged behind tiles with lower refractive index, such that the Cherenkov photons emitted from both layers focus on the photodetection plane. In this scheme, the thick aerogel radiators increase the Cherenkov photon emission, yet the Cherenkov angle resolution is not degraded by emission point uncertainty in the aerogel tile. The appropriate combination of refractive indices is determined by the particle velocity, the distance between the radiator and photodetection plane, and the aerogel thickness. Assuming 2 cm-thick aerogels in each layer, the candidates are $n$ = 1.045 and 1.055 or $n$ = 1.050 and 1.060. The aerogel thickness was constrained by the difficulty of producing large-area, high-refractive-index, crack-free aerogel tiles with thicknesses exceeding 2 cm.

The target transmission length of the A-RICH counter is approximately 40 mm at $\lambda $ = 400 nm. Although aerogel is essentially transparent to visible light, light scattering is non-negligible at shorter wavelengths ($\lambda $ $\sim $400 nm), where most photodetectors exhibit peak quantum efficiency. Only the Cherenkov photons that are not scattered in the aerogel radiator are valuable, because they preserve the Cherenkov angle information. By investigating small aerogel tiles synthesized by the modified KEK method, the approximate transmission lengths of tiles with $n$ = 1.045, 1.050, 1.055, and 1.060 were determined as approximately 45, 40, 35, and 30 mm, respectively \cite{cite2,cite10}. Among these, the transmission length of 30 mm at $n$ = 1.060 was discarded, because transparent downstream aerogels are vital in the focusing radiator scheme. Specifically, all of the emitted Cherenkov photons must pass through the downstream aerogel layer.

High-refractive-index aerogel tiles can also be fabricated by the pin drying technique \cite{cite4}. This method was first explored for producing ultrahigh density aerogel blocks with refractive indices exceeding 1.10. By this technique, extremely transparent aerogel tiles with $n \sim $1.06 were fabricated; i.e., the transmission length was almost doubled \cite{cite11}. Unfortunately, the crack-free yield of large-area pin-dried aerogel tiles never exceeded 50\% in the developmental phase, so the technique was abandoned \cite{cite16}.

The other optical requirements of the radiator are the correct combination of refractive indices in the focusing dual-layer aerogel scheme, and uniform tile density. Assuming that each aerogel tile in the A-RICH counter is 2 cm thick, the refractive indices of the upstream and downstream layers must differ by 0.01. Deviations from the best refractive index combination must not exceed $\pm $0.002; greater deviations will degrade the detector performance (e.g., \cite{citeS9,citeS10}). To ensure superior detector performance over the large-area radiator plane, the uniformity of each monolithic aerogel tile must fulfill a similar condition: $|\delta (n - 1)/(n - 1)| <$ 4\% (i.e., $|\delta \rho /\rho| <$ 4\%).

\subsection{Mechanical requirements}
\label{s3}

The fabricated aerogel tiles should be as large as possible, because the detector performance decreases at the boundaries between adjacent radiators in RICH counters \cite{citeS11}. A similar phenomenon was observed in our test beam experiment \cite{citeS9,citeS12}. Large-area aerogel tiles also increase the detector acceptance. Conversely, the tile size was limited by the capacity of the autoclave in the SCD apparatus. In the present study, the dedicated 140 L autoclave was employed at Mohri Oil Mill, and planned to do the same for the upcoming mass-production. The diameter of the wet gel rack inserted into the autoclave was approximately 41 cm. Therefore, the dimensions of the aerogel tiles demand a trade-off among the sizes of the wet gel rack, the wet-gel tile size that can be safely handled, and the efficiency of mass production.

Cracks in the aerogel tile must be avoided at all costs, for several reasons. First, they cause unwanted scattering of Cherenkov photons in the aerogel tile, losing the Cherenkov angle information. Second, cracked tiles might disintegrate along the crack plane after water jet machining. Finally, cracks reduce the mechanical strength of the tiles, compromising the safety of radiator assembly and installation.




\bibliographystyle{model1-num-names}



\end{document}